\documentstyle[prl,aps,epsfig,floats,twocolumn]{revtex}
\begin{document}

\wideabs{
\title{A polarized-neutron scattering study of the Cooper-pair moment in 
Sr$_{2}$RuO$_{4}$ }

\author{J. A. Duffy$^{1}$, S. M. Hayden$^{1}$, Y. Maeno$^{2,3}$,
Z. Mao$^{2,3}$, J. Kulda$^{4}$ and G. J. McIntyre$^{4}$}

\address{
$^{1}$H. H. Wills Physics Laboratory, University of Bristol, Tyndall 
Avenue, Bristol BS8 1TL, United Kingdom \\
$^{2}$Department of Physics, Graduate School of Science, Kyoto University,
Kyoto 606-8502, Japan \\
$^{3}$CREST, Japan Science and Technology Corporation, Kawaguchi, Saitama 332-0012, Japan \\
$^{4}$Institut Laue-Langevin, B.P. 156, 38042 Grenoble Cedex 9, France
}
\date{14 August 2000}
\maketitle

\begin{abstract}
We report a study of the magnetization density in the mixed state
of the unconventional superconductor Sr$_{2}$RuO$_{4}$.
On entering the superconducting state we find no change in the
magnitude or distribution of the induced moment for a magnetic field
of 1 Tesla applied within the RuO$_2$ planes.  
Our results are consistent with a 
spin-triplet Cooper pairing with spins lying in the basal plane.
This is in contrast  
with similar experiments 
performed on conventional and high-T$_c$ superconductors.

\end{abstract}
}
\narrowtext
Sr$_{2}$RuO$_{4}$ has attracted attention since it was
discovered\cite{Maeno94} to be a superconductor.  The
superconductivity of this compound is interesting because it is
isostructural with the high-$T_c$ material La$_{2-x}$Sr$_{x}$CuO$_{4}$
and because the superconducting state appears unconventional (i.e.\ not
of the $s$-wave singlet type).  The low-temperature normal-state of
Sr$_{2}$RuO$_{4}$ is a quasi-2D Fermi liquid with enhanced
quasiparticles\cite{Mackenzie96}. Soon after the discovery of
superconductivity in Sr$_{2}$RuO$_{4}$, it was
suggested\cite{Rice95,Baskaran96} that the superconducting state might
be unconventional.  This suggestion is
now supported by a number of experiments including: the observation of
a very strong dependence of $T_c$ on impurities\cite{Mackenzie98}, a
temperature-independent $^{17}$O Knight shift on entering the
superconducting state\cite{Ishida98}, a muon-spin rotation study
indicating broken time-reversal symmetry\cite{Luke98}, Andreev
reflection\cite{Laube00}, and the observation of power law $T$-dependences in 
the superconducting state for the electronic heat capacity\cite{Nishizaki00} 
and NQR $T^{-1}_{1}$\cite{Ishida00}.

While there is general agreement that the superconducting state of 
Sr$_{2}$RuO$_{4}$ is unconventional, the nature of the wavefunction
is still controversial
\cite{Rice95,Baskaran96,Sigrist96,Machida96,Agterberg97,Hasegawa00}. A 
knowledge of the 
spin susceptibility in the superconducting state
provides constraints on the pairing wavefunction of a superconductor.  
Such information can be obtained indirectly by nuclear-resonance techniques 
through the measurement of the polarization of the $s$ electrons on a 
given site.
Alternatively, neutron scattering can directly measure the
magnetization density induced by an applied magnetic field.  This
technique was first used by Shull and Wedgewood\cite{Shull66} to study
V$_{3}$Si and more recently it has been applied to
heavy-fermion\cite{Stassis86} and high-$T_c$\cite{Boucherle93}
superconductors.  In this letter we report a study of the induced
magnetization of Sr$_{2}$RuO$_{4}$ through the superconducting
transition.  On entering the superconducting state we find no change
in the magnitude or distribution of the induced moment.  Our results
are in contrast to similar observations on conventional and high-$T_c$
superconductors and are consistent with triplet spin-pairing in
Sr$_{2}$RuO$_{4}$.

The single crystal of Sr$_{2}$RuO$_{4}$ used in this study (C117) was
prepared by a floating-zone method \cite{Mao00b} in an infrared image 
furnace.   A piece of approximate dimensions 
1.5 mm$\times$ 2mm $\times$ 5 mm was cut using a diamond saw.  
A.c. susceptibility measurements indicated a 
sharp superconducting transition with $T_c$=1.47 K and  
$B_{c2}(T=100 \text{mK})$=1.43 T for ${\bf B}$ $\parallel$ 
$[1\overline{1}0]$.  
To perform our neutron scattering experiments, the sample was
glued to a copper stage using Stycast 2850FT. The copper support was
connected to a dilution refrigerator via two 1mm$^2$ diameter copper wires.  
In the present experiment we applied the magnetic field along the 
$[1\overline{1}0]$
direction. The large anisotropy in $B_{c2}$ means that the mutual alignment
of the magnetic field and $[1\overline{1}0]$ is crucial.  
Accurate alignment was
achieved by mounting the sample and copper stage on a micro-goniometer
inside a 2.5 T magnet.  In order to verify that the crystal was
correctly aligned and at low temperature, we performed an in-situ
a.c. susceptibility measurement by mounting two co-axially wound coils
near the sample but out of the neutron beam.

The sensitivity of the present magnetic-moment measurement  can 
be dramatically increased by the use of polarized neutrons.  
Our measurements were performed on the IN20 spectrometer at the 
Institut Laue-Langevin, Grenoble.  A beam of neutrons with a polarization 
greater than 93\%  and with energy $E_i$= 34.8 meV was produced 
using the (111) Bragg reflection of a Heusler monochromator.  We searched 
for possible depolarization of the beam caused by the presence of the vortex
lattice by measuring the polarization of neutrons scattered by the (002)
Bragg reflection using a Heusler analyzer. However, the presence of the 
vortex lattice produced
no detectable depolarization for fields B $\ge$ 1~T under the experimental 
conditions used.

In the present experiment we measure the magnetization density 
${\bf M}({\bf r})$ in the presence of an applied magnetic field ${\bf B}$.  
Because of the periodicity of the crystal, an applied magnetic field induces
a magnetization density with spatial Fourier components  
${\bf M}({\bf G})$ corresponding to reciprocal lattice vectors
${\bf G}$, where
\begin{eqnarray}
\label{eq_M}
{\bf M}({\bf r}) & = & \frac{1}{\nu_{0}}\sum_{{\bf G}} 
{\bf M}({\bf G}) \exp(-i{\bf G} \cdot {\bf r})
\end{eqnarray}
and $\nu_{0}$ is the unit-cell volume.
The Fourier components of the magnetization density are given by 
\begin{eqnarray}
\label{eq_F} 
{\bf M}({\bf G}) & = & \int_{\text{cell}} {\bf M}({\bf r}) 
\exp(i {\bf G} \cdot{\bf r}) d{\bf r}.
\end{eqnarray}
A diffraction experiment allows these spatially varying 
components of the magnetization to be measured, even in the superconducting 
state.

Neutrons interact with condensed matter both through the 
strong nuclear
interaction and through the electromagnetic interaction. If the
neutron momentum transfer $\bbox{\kappa}={\bf k}_i-{\bf k}_f$ equals a
reciprocal lattice vector ${\bf G}$, i.e.\ we satisfy the Bragg
condition, then scattering occurs both because of the periodicity of the
nuclear density and because of the microscopic periodicity of the 
magnetization density.  The two scattered waves interfere. 
For neutrons with initial and final
spin polarizations $\bbox{\sigma}_{i}$ and $\bbox{\sigma}_{f}$, the 
total cross section 
is\cite{Squires78,Brown92},
\begin{equation}
\label{eq1}
\frac{d \sigma}{d \Omega}_{\sigma_{i} \rightarrow \sigma_{f}} 
 \propto  
\left| \langle \sigma_{i} | 
\frac{\gamma r_{0}}{2 \mu_{B}}  
\bbox{\sigma} \cdot \bbox{\hat{\kappa}} \times 
\left\{ {\bf M}(\bbox{\kappa}) \times 
\bbox{\hat{\kappa}} \right\} + 
F_{N}(\bbox{\kappa})
| \sigma_{f} \rangle \right|^{2}
\end{equation}
where $\gamma r_0$= 5.36$\times 10^{-15}$~m, $F_{N}(\bbox{\kappa})$ is the
nuclear structure factor (in units of m~f.u.$^{-1}$) and
${\bf M}(\bbox{\kappa})$ has units of $\mu_{B}$~f.u.$^{-1}$.
The sign of the magnetic term in Eq.~\ref{eq1} is controlled by the 
polarization of the neutrons.  By reversing
the polarization of the incident neutrons we are able to isolate
the term due to the interference between magnetic and nuclear contributions, 
yielding the magnetization density.  Experimentally
we measure the flipping ratio $R$, defined as the ratio of cross sections
for initial neutron-spin states which are parallel or antiparallel to the 
applied magnetic field and with arbitrary final spin state.  
In the present experiment, the applied field, and hence neutron polarization,
is perpendicular to the scattering vector $\bbox{\kappa}$.  Under these
circumstances the flipping ratio $R$ evaluated from Eq.~\ref{eq1} is only 
sensitive to the component of magnetization 
parallel to the
applied field, $M_{\parallel}(\bbox{\kappa})$.  Because the induced moment 
is small, the experiment is carried out in the limit
$(\gamma r_{0}/2 \mu_{B}) |{\bf M}(\bbox{\kappa})|
/|F_N(\bbox{\kappa})| \approx 0.001 \ll 1$. In this limit, the flipping ratio 
derived from Eq.~\ref{eq1} is,
\begin{equation}
\label{eq3}
R  =  \frac{|F_{N}(\bbox{\kappa})-
(\gamma r_{0} /2 \mu_{B}) {M_{\parallel}}(\bbox{\kappa})|^{2}}
{|F_{N}(\bbox{\kappa})+
(\gamma r_{0}/2 \mu_{B}) {M_{\parallel}}(\bbox{\kappa})|^{2}} 
 \approx  
1 - \frac{2 \gamma r_{0}}{\mu_{B}}
\frac{{M_{\parallel}}(\bbox{\kappa})}
{F_{N}(\bbox{\kappa})}. 
\end{equation}
%
\begin{figure}[t]
\centering
\epsfxsize=8.0cm
\epsffile{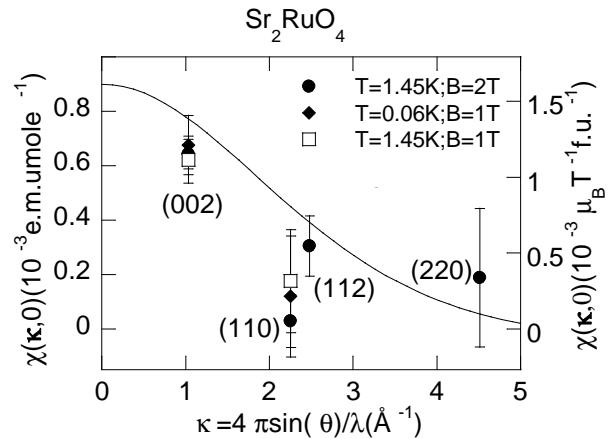}
\caption{ 
The wavevector-dependent susceptibility $\chi(\bbox{\kappa},0)$ determined 
from the induced moment for applied magnetic fields of 1 and 2 Tesla applied 
along the $[1\overline{1}0]$ direction.  Solid line is a scaled Ru form factor
\protect\cite{Brown92}
}
\label{chiqq}
\end{figure}
As the nuclear structure factors $F_{N}(\bbox{\kappa})$ are known from the 
crystal structure, Eq.~\ref{eq3} directly gives the magnetization
${\bf M}({\bf G})$. 
Fig.~\ref{chiqq} shows the susceptibility 
$\chi(\bbox{\kappa},0)={\bf M}(\bbox{\kappa})/B$ for a number of
wavevectors $\bbox{\kappa}$ determined from
the measured flipping ratios $R$ and converted into magnetic moment using
Eq.~\ref{eq3}. The measured flipping ratios were corrected for
imperfect beam polarization and  finite flipper efficiency. Other possible 
corrections including 
extinction, absorption, incoherent scattering, the neutron spin-orbit 
interaction \cite{Squires78} , half-wavelength contamination in 
the incident beam were estimated to be small.  In order to allow us to 
enter the mixed state, 
measurements were generally made with an applied field of 1 Tesla.  However,
Fig.~\ref{chiqq} shows that measurements in the normal state at 
$\bbox{\kappa}=(002)$ and $B$=2 Tesla yielded consistent results. 

Neutron scattering measures the total moment, thus Fig.~\ref{chiqq} 
includes contributions from diamagnetic, orbital and spin components of the 
susceptibility.  The fall off of the susceptibility 
$\chi(\bbox{\kappa},0)$ with increasing 
$|\bbox{\kappa}|$ is due to the finite extent 
of the induced moment in space.
Fig.~\ref{chiqq} shows the Ru form factor (solid line) 
\cite{Brown92} scaled to the measured bulk susceptibility \cite{Maeno97} of 
Sr$_{2}$RuO$_{4}$, $\chi_{ab}$=0.9$\times 10^{-3}$ e.m.u. mole$^{-1}$.
The poor agreement of the $\bbox{\kappa}$=(110) 
component with the Ru form factor may suggest the presence of a
significant induced moment on the oxygen atoms.     

\begin{figure*}[t]
\centering
\epsfxsize=18.0cm
\epsfbox{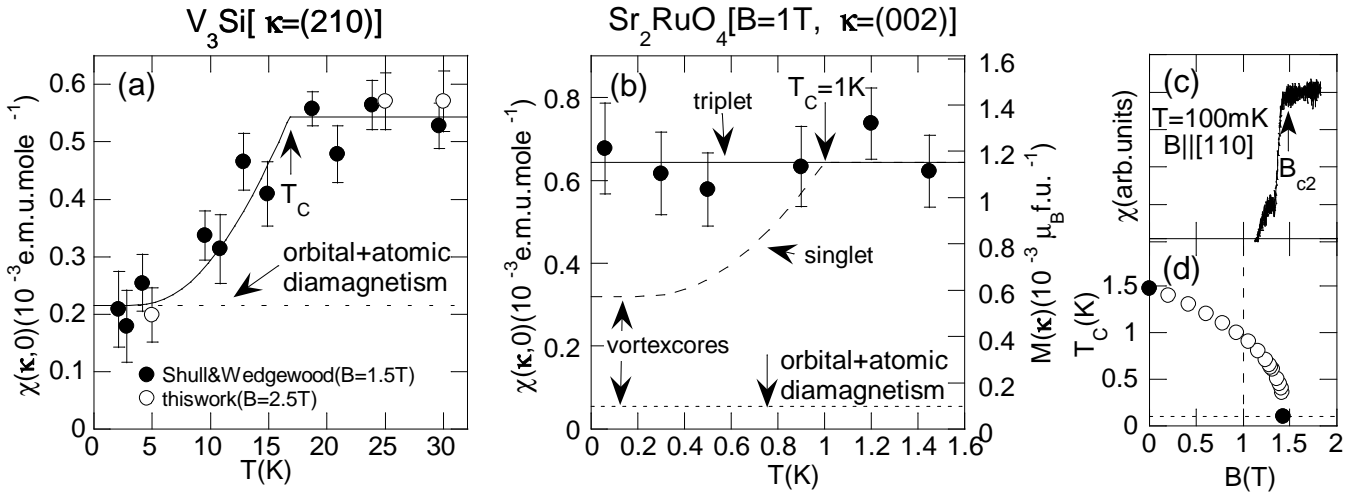}
\caption{ 
(a) The susceptibility of the conventional $s$-wave singlet superconductor
V$_3$Si measured by Shull and Wedgewood \protect\cite{Shull66} 
using the present neutron scattering method.  At low
temperatures a residual orbital contribution to the susceptibility remains.
(b) The temperature dependence of the susceptibility and induced moment 
of Sr$_2$RuO$_4$ measured using polarized neutron scattering.  The dashed
line is the Yosida \protect\cite{Yosida58} behavior expected for a 
singlet-paired superconductor. 
Counting times were 8 hours per spin polarization.
(c) In-situ a.c. susceptibility measurement to demonstrate accurate alignment
of the sample. (d) The $B_{c2}(T)$ for Sr$_2$RuO$_4$ for ${\bf B}
\parallel$ 
$[1\overline{1}0]$.  Closed circles are the present measurements; open circles are 
from Ref.~\protect\onlinecite{Mao00a}.
The horizontal dashed line denoted the trajectory of the in-situ 
susceptibility measurement in panel (c) and the vertical line is the
trajectory of the temperature-dependent susceptibility measurement in (b).
}
\label{chiT}
\end{figure*}

Before discussing our results in the superconducting state of 
Sr$_{2}$RuO$_{4}$, we will briefly discuss the same measurement of
the spin susceptibility in the conventional superconductor V$_3$Si.
In a conventional superconductor with spin-singlet pairing,
the spin susceptibility is suppressed on entering the superconducting
state because electrons with anti-parallel spins pair up.  For
$B \ll B_{c2}$ the temperate dependence is described by the 
Yosida function \cite{Yosida58}. Wedgewood and 
Shull \cite{Shull66} performed a polarized-neutron measurement of the
susceptibility
on the conventional superconductor, V$_{3}$Si, and observed
a reduction of the susceptibility, due to the formation of spin
singlets ($S=0$) [see Fig.~\ref{chiT}(a)]. We have reproduced the 
Wedgewood-Shull result using the same experimental set-up as for our
Sr$_{2}$RuO$_{4}$ measurements.  Our results are consistent with Wedgewood 
and Shull
and are shown as open circles in Fig.~\ref{chiT}(a). 
%

Having observed the induced moment in the normal state 
of Sr$_{2}$RuO$_{4}$,  we proceeded to 
investigate the effect of the superconductivity.  Because of the low
$T_c$ and strongly anisotropic $B_{c2}$ it was important to verify that
the sample was in good thermal contact with the dilution refrigerator and
well aligned with the applied magnetic field. 
Fig.~\ref{chiT}(c) shows an in-situ measurement of the a.c. susceptibility, 
made using the balance output of an inductance bridge, plotted against 
applied field $B$ for T=100 mK.  The kink corresponds to $B_{c2}$=1.43 Tesla,
which is indistinguishable from the published value \cite{Mao00a}, 
demonstrating that the sample was well-aligned.   
In order to ensure penetration of the magnetic field throughout the sample
the sample was always ``field-cooled''.

Fig.~\ref{chiT}(b) shows the temperature dependence of the induced 
moment corresponding to the (002) Bragg position for $B$=1 Tesla. This 
component was chosen for detailed study because its amplitude is proportional 
to the sum of the moments induced on the in-plane ruthenium and oxygen atoms 
$(m_{\text{Ru}}+2m_{\text{O}})$.  For a 1 Tesla field applied parallel to 
$[1\overline{1}0]$, $T_c =$1 K (See Fig.~\ref{chiT}(d)).  On entering the superconducting 
state, we find that there is no change in this component of the induced moment
within the experimental error. 
In contrast to the V$_3$Si measurement, we investigated Sr$_2$RuO$_4$ at
relatively high fields, $B/B_{c2}=0.68$, thus the presence of normal 
vortices leads to a significant density of quasiparticles and finite 
spin susceptibility in the mixed state.  
Using the measured linear heat capacity in the
superconducting state\cite{Nishizaki00} we estimate\cite{note} the 
zero-temperature spin susceptibility 
$\chi(T \rightarrow 0, B=1T)=0.45\chi_{\text{normal}}$. 
The dashed line in Fig.~\ref{chiT}(b) is a Yosida function modified to include
the finite susceptibility in a field: this prediction is still at variance 
with the data. Thus, the absence of a change in spin susceptibility 
is not compatible with spin-singlet or even-parity pairing.    

The absence of a change in the spin susceptibility can be explained if 
Cooper pairs form from electrons with parallel spins.  Such ``equal-spin
pairing'' (ESP) was first proposed in the context of $^{3}$He by Anderson 
and Morel \cite{Anderson84}.  Within an ESP scenario the superconducting 
state is
a superposition of the two possible ($S=1$) parallel paired states.  In an
applied magnetic field, one state is favored yielding a net spin moment
and the same susceptibility as in the normal state \cite{Leggett75}.
An EPS-type pairing implies an odd-parity or spin-triplet state, thus the 
present experiment supports the notion that the superconducting
wavefunction in Sr$_2$RuO$_4$ has an odd-parity representation.  
There are five 
unitary odd-parity representations ($\Gamma^{-}_{1-5}$) of the order 
parameter for the crystal point group $C_{4h}$ \cite{Rice95,Sigrist99}.  
Of the allowed 
representations only the
degenerate $\Gamma^{-}_{5}$ ($E_{u}$) or 
${\bf d}(\hat{\bf k})=\hat{\bf z}(\hat{k}_{x} \pm i \hat{k}_{y})$ 
state is expected to show no change in its 
spin susceptibility for magnetic fields applied in the basal
plane.  

So far we have discussed the spin part of the Cooper-pair wavefunction. 
The $\Gamma^{-}_{5}$ state is special in that the orbital part of the
wavefunction suggests that the paired electrons have relative angular 
motion with orbital angular momentum $L_{z}=\pm 1$.  $\mu$SR measurements 
\cite{Luke98}
reveal a spontaneous internal magnetic field in the superconducting
state which is thought to be associated with the internal orbital moment.
The present experiment is sensitive to the total electronic moment 
(${\bf S+L}$) in the superconducting state.  Under 
the present experimental conditions, we measure magnetic moments 
parallel to the applied field.
The absence of a change in the orbital moment measured by
the present experiment is entirely consistent with the $\Gamma_{5}^{-}$
assignment above. Firstly, because the moment is expected to be parallel
to the $c$-axis and, secondly, because the bulk orbital moment is expected
to be small\cite{Mineev99}.

At first sight our results appear to contradict recent heat capacity 
\cite{Nishizaki00} and other experiments \cite{Ishida00} which suggest that 
the superconducting gap is strongly
anisotropic, possibly having nodes for certain directions.  However, a
strongly anisotropic gap function is still allowed within
the  $\Gamma_{5}^{-}$ representation.  For example, the anisotopic 
$f$-states which
have recently been proposed \cite{Hasegawa00} still have the 
$\Gamma_{5}^{-}$ representation
and are consistent with our interpretation.

Our results complement and contrast with recent NMR measurements 
\cite{Ishida98} of the $^{17}$O
Knight shift in the mixed state of Sr$_{2}$RuO$_{4}$.  These do not 
detect a reduction in the spin susceptibility.   The 
Knight shift measures the polarization of the $s$-electrons at a given
site: electrons in other orbitals and on other sites are probed because of 
the overlap of orbitals. In contrast, the present neutron-scattering 
measurement directly measures the spatially-averaged total moment. 

Measurements of the induced moment in the mixed state have also been
performed on other unconventional superconductors.  In the high-temperature
superconductor YBa$_2$Cu$_3$O$_{\text{6+x}}$ \cite{Boucherle93} a suppression
of the spin susceptibility is observed which is consistent with an even-parity 
or singlet  
pairing.  In contrast, the heavy-fermion superconductors UPt$_3$ and 
UBe$_{13}$ \cite{Stassis86,Tou96} show no reduction in the spin susceptibility 
on entering the superconducting state suggesting that they have odd-parity 
pairing.

In summary, we have used a spin-polarized neutron-scattering technique to
measure the magnetization in the mixed state of the unconventional 
superconductor Sr$_{2}$RuO$_{4}$.  We find that for a 1 Tesla field applied
parallel to the $[1\overline{1}0]$ basal-plane direction, there is no 
detectable change 
in the component of the moment parallel to the applied field.  Our results
strongly support the identification of the paired state of Sr$_2$RuO$_4$ 
as the  $\Gamma^{-}_{5}$ or 
${\bf d}(\hat{\bf k})=\hat{\bf z}(\hat{k}_{x} \pm i \hat{k}_{y})$ state. 

We are grateful to D. Bintley, N. Clayton, and J.-L. Raggazoni for their
help in preparing this experiment and G. Aeppli, J. Annett, E. M. Forgan 
and A. P. Mackenzie for useful discussions.            


\begin{references}
\bibitem{Maeno94}
Y. Maeno {\it et al.}, Nature {\bf 372}, 532 (1994).
\bibitem{Mackenzie96}
A. P. Mackenzie {\it et al.}, Phys. Rev. Lett. {\bf 76}, 3786 (1996).
\bibitem{Rice95}
T. M. Rice and M. Sigrist, J. Phys.: Condens. Matter {\bf 7}, 643 (1995).
\bibitem{Baskaran96}
G. Baskaran, Physica B {\bf 223-224}, 490 (1996).
\bibitem{Mackenzie98}
A. P. Mackenzie {\it et al.}, Phys. Rev. Lett. {\bf 80}, 161 (1998).
\bibitem{Ishida98}
K. Ishida {\it et al.}, Nature {\bf 396}, 658 (1998).
\bibitem{Luke98}
G. M. Luke {\it et al.}, Nature {\bf 394}, 558 (1998).
\bibitem{Laube00}
F. Laube {\it et al.}, Phys. Rev. Lett. {\bf 84}, 1595 (2000).
\bibitem{Nishizaki00}
S. Nishizaki, Y. Maeno, and Z. Mao, J. Phys. Soc. Japan {\bf 69}, 572 (2000).
\bibitem{Ishida00}
K. Ishida {\it et al.}, Phys. Rev. Lett. {\bf 84}, 5387 (2000).
\bibitem{Sigrist96}
M. Sigrist and M. Zhitomirsky, J. Phys. Soc. Japan {\bf 65}, 3452 (1996).
\bibitem{Machida96}
K. Machida, M. Ozaki, and T. Ohmi, J. Phys. Soc. Japan {\bf 65}, 3720 (1996).
\bibitem{Agterberg97}
D. F. Agterberg, T. M. Rice, and M. Sigrist, Phys. Rev. Lett.
{\bf 78}, 3374 (1997).
\bibitem{Hasegawa00}
Y. Hasegawa, K Machida, and M. Ozaki, J. Phys. Soc. Japan 
{\bf 69}, 336 (2000).
\bibitem{Shull66}
C. G. Shull and F. A. Wedgewood, Phys. Rev. Lett. {\bf 16}, 513 (1966).
See also C. G. Shull and  R. P. Ferrier, ibid {\bf 10}, 295 (1963) and 
C. G. Shull, ibid {\bf 10}, 297 (1963).
\bibitem{Stassis86}
C. Stassis {\it et al.}, Phys. Rev. B {\bf 34} 4382 (1986).
\bibitem{Boucherle93}
J. X. Boucherle {\it et al.}, Physica B {\bf 192}, 25 (1993).
\bibitem{Mao00b} 
Z.Q. Mao, Y. Maeno, and H. Fukazawa, {\it Crystal growth of
Sr$_2$RuO$_4$}, to be published in Mat. Res. Bull. {\bf 35} (2000) No. 11. 
\bibitem{Squires78}
G. L. Squires, {\it Introduction to the Theory of Thermal Neutron Scattering} 
(Cambridge, 1978).
\bibitem{Brown92}
P. J. Brown in {\it International Tables for Crystallography, vol C} 
(Kluwer, Bordrecht, 1992) p391.
\bibitem{Maeno97} 
Y. Maeno {\it et al.} J. Phys. Soc. Jap. {\bf 66} 1405 (1997).
\bibitem{Yosida58} 
K. Yosida, Phys. Rev. {\bf 110}, 769 (1958).
\bibitem{Mao00a}
Z. Q. Mao {\it et al.}, Phys. Rev. Lett. {\bf 84} 991 (2000).
\bibitem{note}
In calculating $\chi_{\text{normal}}$, we assume 
$\chi(B)/\chi_{\text{normal}}=\gamma(B)/\gamma_{\text{normal}}$, 
where $\gamma$ is
the linear coefficient of the electronic heat capacity. 
\bibitem{Anderson84}
See e.g. P. W. Anderson {\it Basic Notions of Condensed Matter Physics} 
(Benjamin/Cummings, 1984) p327.
\bibitem{Leggett75}
In the analogous transition from the normal to the superfluid A phase 
of $^{3}$He, a small increase of the spin susceptibility is expected
theoretically, and found experimentally. See A. J. Leggett, Rev. Mod. Phys.
{\bf 47}, 404 (1975) and J. C. Wheatley, ibid {\bf 47}, 423 (1975) 
respectively.  Such a change appears to be beyond the resolution of the
present experiment.  
\bibitem{Sigrist99}
M. Sigrist {\it et al.}, cond-mat/9902214
\bibitem{Mineev99}
V. P. Mineev and K. V. Samokhin, 
{\it Introduction to Unconventional Superconductivity} 
 (Gordon and Breach, 1999) p106.
\bibitem{Tou96}
H. Tou {\it et al.}, Phys. Rev. Lett. {\bf 77}, 1374 (1996).
\end{references}
\end{document}